\documentclass[journal]{IEEEtran}

\usepackage{graphicx}
\usepackage{amsmath}
\usepackage{amssymb}
\usepackage{cite}
\usepackage{url}
\usepackage[hidelinks]{hyperref}
\usepackage{placeins}

\begin{document}

\title{PersianVox: A Prosody-Aware Approach for Speech Dataset Generation from In-the-Wild Data}

\author{Saeedreza Zouashkiani,
        Soheil Khalesi,
        Saman Soleimani Roudi,
        Sajjad Amini,
        and Shahrokh Ghaemmaghami%
\thanks{The authors are with the Electronics Research Institute,
Sharif University of Technology, Tehran, Iran.}%
\thanks{Corresponding author: Sajjad Amini
(e-mail: s\_amini@sharif.edu).}%
\thanks{Demo available at:
\url{https://saeedzou.github.io/persianvox-demo}.}}

\maketitle

\begin{abstract}
Advancement of zero-shot text-to-speech synthesis is currently hindered for low-resource languages by the scarcity of large-scale, high-fidelity speech datasets. Traditional alignment-based methods require rare verbatim transcripts, while standard in-the-wild pipelines often rely on single-model automatic speech recognition and silence-based segmentation, leading to transcription errors and truncated prosody. To address these challenges for the Persian language, this paper introduces PersianVox, a fully automated pipeline designed to generate high-quality speech corpora from unlabeled web data. Our approach integrates a novel prosody-aware segmentation strategy that utilizes acoustic turn-detection to preserve linguistic completeness and optimize utterance duration for long-context modeling. Furthermore, we employ a dual-model agreement mechanism, leveraging two distinct model architectures to filter unreliable transcriptions without ground truth. This pipeline yields a 2,400-hour multi-speaker dataset, the largest open-source speech resource available for Persian to date. Additionally, we provide the first comparative benchmark of speech quality assessment methods for Persian, releasing a human-annotated subset to facilitate future research.
\end{abstract}


\IEEEpeerreviewmaketitle

\section{Introduction}
	\IEEEPARstart{R}{ecent} advancements in text-to-speech (TTS) synthesis rely heavily on large-scale, high-quality datasets curated via alignment \cite{Kurzinger2020Jul, rastorgueva2023nemo} or in-the-wild automatic speech recognition (ASR) transcriptions \cite{He2025Jan}. However, both methodologies present significant bottlenecks for low-resource languages. Alignment-based approaches are limited by the scarcity of clean, paired text-audio resources, while in-the-wild pipelines depend on highly accurate ASR models that are often unavailable for underrepresented languages. Furthermore, these in-the-wild pipelines typically employ silence-based segmentation, which frequently truncates utterances mid-sentence and disrupts the long-range prosody crucial for modern TTS modeling. Consequently, without high-fidelity transcriptions, robust ASR, and intelligent segmentation, neither approach offers a scalable path for low-resource dataset creation.
	
	Persian exemplifies these challenges, as existing resources are either limited in scale \cite{Qharabagh2024Sep} or restricted by manual curation and copyright limitations \cite{Adibian2023Aug}. To address these limitations, we propose a fully automated pipeline based on turn-aware segmentation and dual-ASR agreement, enabling scalable construction of high-quality Persian speech data. Applying this framework, we curate PersianVox, a 2,400-hour multi-speaker dataset, and make the following contributions: (1) We propose a 
    prosody-aware segmentation strategy that utilizes acoustic turn-detection 
    to achieve linguistic completeness and optimize duration for 
    long-context TTS, in contrast to the silence-based voice activity detection (VAD) segmentation used 
    by pipelines such as Emilia and AutoPrep, which frequently truncates 
    utterances mid-sentence. (2) We introduce a dual-ASR agreement framework 
    that filters unreliable transcriptions without ground truth, offering a 
    scalable alternative to single-ASR pipelines for low-resource languages. 
    (3) We provide a comparative analysis of speech quality assessment (SQA) methods for Persian, 
    releasing a 1.6-hour human-annotated mean opinion score (MOS) subset, the first such resource 
    for the language\footnote{\url{https://huggingface.co/datasets/saeedzou/persianvox_mos}}. (4) We demonstrate the effectiveness of our approach by 
    releasing the largest open-source, multi-speaker Persian dataset to date\footnote{\url{https://huggingface.co/datasets/saeedzou/persianvox}}.
	
\section{Related Work}
    A common way to build datasets is to split long audio recordings into shorter, utterance-level clips using their matching text transcripts, a process that can be automated via CTC-based automatic alignment. Kürzinger et al. \cite{Kurzinger2020Jul} proposed CTC-segmentation, which uses a pre-trained ASR model with CTC loss to align text and audio automatically. Instead of relying on manually specified timestamps like traditional forced alignment, it computes frame-level character probabilities to find utterance start/end times, and remains robust even when the audio has extra untranscribed speech at the start or end.

	The main drawback of alignment methods is that they need a verbatim transcript for the audio. Producing such high-fidelity transcripts is labor-intensive, and the limited amount of existing data with clean, matching text caps how scalable this approach can be. To address this, pipelines have been built to handle "in-the-wild" audio with no labels at all. Frameworks like AutoPrep \cite{Yu2023Sep} and Emilia-Pipe \cite{He2025Jan} automatically turn unlabeled audio into labeled datasets by running it through steps such as background noise and music removal, speaker diarization, and voice activity detection (VAD), then using a strong ASR model to produce ground-truth transcriptions. As a final quality-control step, pipelines like AutoPrep and Emilia filter out low-quality segments with a DNSMOS (Deep Noise Suppression Mean Opinion Score) threshold. However, our experiments found that DNSMOS generalizes poorly to Persian, frequently giving low scores to high-quality audio, which makes it an unreliable filter. To get past this bottleneck, we explore more robust, representation-based SQA methods like SCOREQ \cite{ragano2025scoreqspeechqualityassessment}, which use contrastive learning to better capture the perceptual quality manifold in out-of-distribution languages. Table~\ref{tab:persian_datasets} summarizes existing Persian datasets, highlighting the gap in scale and openness that PersianVox fills.
    
    \begin{table}[!t]
      \centering
      \caption{Existing Persian datasets for TTS synthesis.}
      \label{tab:persian_datasets}
      \begin{tabular}{lccc}
        \hline
        \textbf{Dataset} & \textbf{Speakers} & \textbf{Hours} & \textbf{License (Availability)} \\ \hline
        ManaTTS \cite{Qharabagh2024Sep}          & 1     & $\sim$86 & CC0 1.0 (\checkmark)   \\
        DeepMine Multi-TTS \cite{Adibian2023Aug} & 67    & 120      & Unknown (on request)   \\
        PersianVox (ours)                        & 3,248 & 2,400    & CC-BY-4.0 (\checkmark) \\ \hline
      \end{tabular}
    \end{table}
    
	\section{Method}
\begin{figure}[!t]
            \centering
            \includegraphics[width=0.45\textwidth]{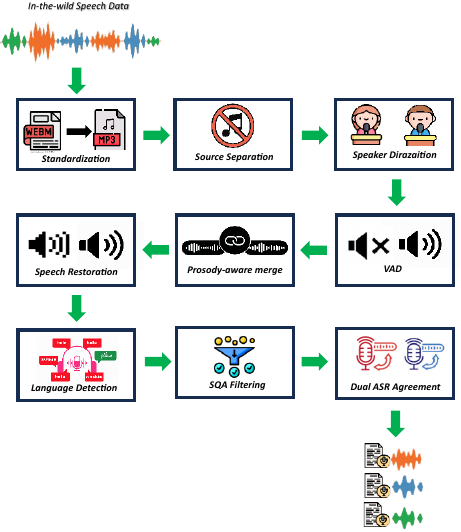}
            \caption{Overall pipeline structure.}
            \label{fig:pipeline}
        \end{figure}

	As illustrated in Fig.~\ref{fig:pipeline}, our proposed approach transforms raw, in-the-wild audio into a high-quality dataset through a systematic, multi-stage process. Since the proposed pipeline consists of sequential stages without an explicit feedback mechanism, errors introduced at an early stage may propagate to subsequent stages. For example, source-separation artifacts may affect diarization and segmentation, while inaccurate segment boundaries or restoration artifacts may reduce the reliability of language identification, quality assessment, and ASR transcription. We mitigate these cascading effects through input standardization, confidence-based filtering, boundary checks, and the final dual-ASR agreement stage. This section describes each pipeline stage in detail: preprocessing, utterance segmentation, speech restoration, utterance filtering, and dual-ASR agreement.
	
	\subsection{Preprocessing}
    	In-the-wild audio collected from online sources often suffers from inconsistent formats, background music overlays, and varying loudness levels, all of which hinder downstream processing such as VAD and ASR. Therefore, we apply a rigorous standardization and cleaning pipeline to ensure consistency and suppress interference from non-speech signals \cite{He2025Jan}.
	
	\subsubsection{Initial Language Detection} \label{subsubsec:initial_lang}
    	To minimize computational overhead on diverse raw audio, we perform early-stage filtering by analyzing 10 random, speech-dominant segments (root-mean-square (RMS) amplitude $>0.01$) per recording. Using Whisper Large v3, we determine the language via maximum voting. Recordings are retained only if the average target language probability exceeds 90\%, establishing our initial Persian-dominant base dataset.

	\subsubsection{Standardization} 
    	We standardize raw audio by converting it to mono, 16-bit PCM WAV at 24 kHz. Audio is normalized to $-20$ dBFS via peak normalization with a $\pm 3$ dB gain cap, and samples are scaled to $[-1,1]$. This ensures consistent signal characteristics for segmentation and inference.
	
	\subsubsection{Music Background Removal} 
    	We suppress background music using neural source separation with the open-source Ultimate Vocal Remover model (UVR-MDX-Net Inst 33 \cite{kim2021kuielab}. The model uses spectral masking for vocal isolation and extracts cleaner speech tracks by reducing musical overlays in podcasts and multimedia sources.
	
	\subsection{Utterance Segmentation}
    	To convert long-form recordings into clean, single-speaker utterances suitable for dataset construction, we employ a two-stage segmentation process involving speaker diarization followed by fine-grained prosody-aware segmentation.
	
	\subsubsection{Speaker Diarization} 
    	We first apply speaker diarization using PyAnnote's community version \cite{Bredin23}, a state-of-the-art toolkit based on neural speaker embeddings and agglomerative clustering. This step segments each recording into intervals of homogeneous speaker identity.
	
	\subsubsection{Prosody-Aware Segmentation} \label{subsubsec:prosody_aware}
    	Standard VAD-based pipelines (e.g., Emilia) split audio at any silence exceeding a brief threshold (e.g., 2 s), producing a power-law duration distribution dominated by short ($<5$ s), mid-sentence clips that impair a TTS model's ability to learn long-range prosodic dependencies \cite{makarov2022simpleeffectivemultisentencetts}.
        We instead propose a prosody-aware, target-driven strategy. After applying Pyannote Segmentation 3.0 \cite{Bredin23} to detect speech regions, we use \textit{Smart Turn v3.1} \cite{smartturnv3}, an acoustic turn-detection model that achieves 80\% accuracy on our annotated Persian subset, to evaluate the linguistic completeness of candidate boundaries. A dynamic merging algorithm then samples target durations from a Gaussian distribution ($\mu=12~\mathrm{s}$, $\sigma=4~\mathrm{s}$). This range was chosen empirically to match the practical operating regime of current neural TTS systems: utterances must be long enough to provide phrase- and sentence-level context for prosody modeling, yet not so long that alignment complexity, memory demands, and training instability increase. The algorithm iteratively merges adjacent same-speaker segments across short gaps until reaching either the target length or a confident complete boundary, subject to a hard 30-sec cap. This shifts the duration distribution away from truncated VAD bursts toward a more balanced range, giving TTS models the extended context needed for superior prosodic realization.
	
	\subsection{Speech Restoration}
    	Beyond background music, most in-the-wild recordings suffer from low bandwidth and residual degradations such as reverberation, codec distortion, and environmental noise that persist after source separation. To address this, we introduce a dedicated speech restoration stage using SIDON \cite{nakata2026sidonfastrobustopensource}, an open-source multilingual model that converts degraded speech into studio-quality audio via a parametric resynthesis framework built on a w2v-BERT 2.0-based feature predictor and a HiFi-GAN-style vocoder \cite{kong2020hifigangenerativeadversarialnetworks}. SIDON is applied to all utterances passing preprocessing, operating on the music-suppressed tracks from UVR-MDX-Net, to eliminate residual artifacts before utterances enter the filtering and dual-ASR stages.        

	\subsection{Utterance Filtering}
    	To ensure that only clean and relevant content in the target language is included in the final dataset, we apply both language identification and acoustic quality filtering.
	
	\subsubsection{Language Identification} 
    	Each utterance is transcribed using Whisper Large v3, which includes a built-in language identification module. We retain only those utterances where target language is detected with high confidence (probability $>$ 0.95). This step is essential for excluding non-Persian or code-switched content, which is common in web-crawled or podcast audio.
	
	\subsubsection{Acoustic Filtering} 
    	We employ a neural SQA-based filtering strategy. Based on our evaluations (detailed in Section~\ref{section:sqa_benchmark}), SCOREQ serves as the most reliable predictor for Persian speech quality. We retain only those utterances that exceed a specific SCOREQ threshold, ensuring a high level of perceptual quality across the final dataset.
    	
    	Together, these filters ensure that the resulting dataset is composed of high-quality Persian utterances, both in terms of linguistic fidelity and acoustic clarity.
	
	\subsection{Dual-ASR Agreement}
        This design is motivated by two key observations from prior work. First, CTC and RNN-T models produce complementary and largely uncorrelated error patterns, making their agreement a more informative indicator of transcription correctness than the output of either model alone \cite{doutre21_interspeech, gitman23_interspeech}. Second, large-scale data-curation pipelines use cross-model transcript consensus as a substitute for ground-truth references to identify reliable transcriptions \cite{bhogale24_interspeech}. To filter unreliable transcriptions in the absence of ground truth, we adopt a dual-ASR agreement strategy, that leverages two proprietary models with distinct architectures, to minimize correlated errors. These models are a 114M-parameter FastConformer Hybrid (CTC head), and a 600M-parameter Parakeet RNN-T. Their differing decoding mechanisms yield complementary error patterns. For instance, CTC tends to produce alignment-related deletions, while RNN-T is more prone to substitutions or insertions arising from label prediction. Consequently, agreement between their outputs serves as a stronger indicator of transcription reliability, especially for noisy or low-resource data. In such settings, consensus between architecturally distinct models becomes a more discriminative filter for identifying reliable transcriptions. We compute the character error rate (CER) and word error rate (WER) between their outputs, retaining only utterances that satisfy strict consistency thresholds (CER $<12.5\%$ and WER $<15\%$), determined empirically to balance transcription fidelity against dataset yield. We additionally enforce boundary precision by discarding segments with high ``edge CER'' ($>50\%$) on the first and last seven characters, and select the transcription from the larger Parakeet model as the final label for passing utterances.

\section{Dataset Statistics}
	\subsection{Data Collection and Filtering}
        We curated our initial dataset from Creative Commons-licensed Persian-language YouTube content using a two-stage keyword and channel-based search strategy, collecting 9,371.52 hours of raw audio across 32,207 videos from 792 distinct channels. After language filtering, 6,179.66 hours of Persian-dominant recordings were retained. We treat this 6k-hour subset as the raw dataset for all subsequent processing, including preprocessing, segmentation, language and acoustic quality filtering, and dual-ASR agreement (Table~\ref{tab:data_funnel}).
	
    	\begin{table}[!t]
    		\centering
    		\caption{Data reduction statistics throughout the pipeline.}
    		\label{tab:data_funnel}
    		\begin{tabular}{lc}
    			\hline
    			\textbf{Stage} & \textbf{Hours Retained} \\ \hline
    			Raw Audio & 6,179.66 \\
    			Utterance Segmentation & 4,355.86 \\
    			Language Filtering & 4,037.90 \\
    			Quality Filtering & 3,842.96 \\
    			Dual-ASR Agreement (Final) & 2,408.67
    			\\ \hline
    		\end{tabular}
    	\end{table}

    \subsection{Corpus Statistics}
\begin{figure}[!t]
    		\centering
    		\includegraphics[width=0.85\linewidth]{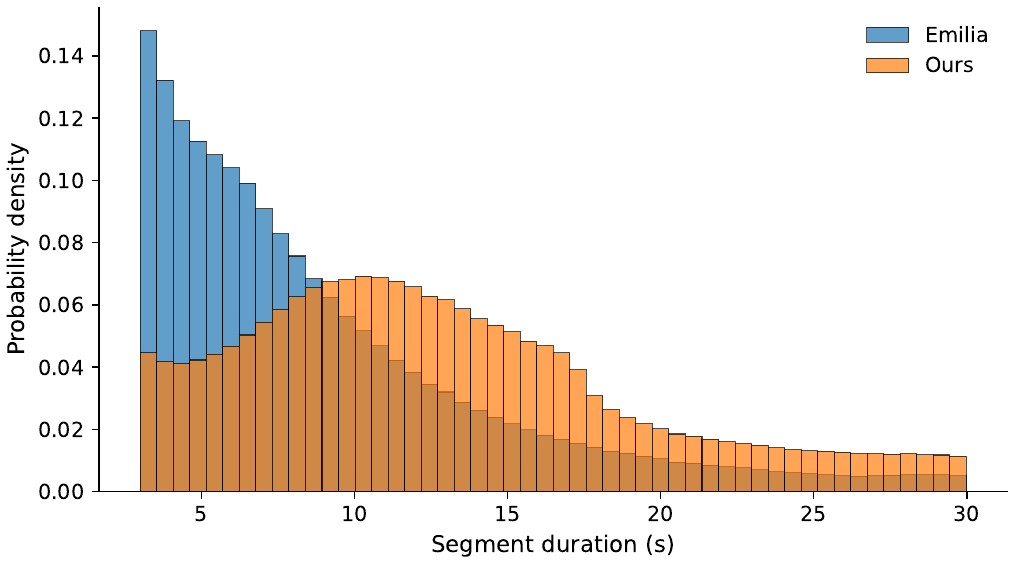}
    		\caption{Utterance duration distributions.}
    		\label{fig:duration_distribution}
    	\end{figure}

        Table~\ref{tab:corpus_stats} summarizes key statistics of the final 
        PersianVox corpus. The dataset comprises 625,192 utterances totaling 
        2,408.67 hours, contributed by an estimated 3,248 unique speakers, 
        obtained by globally linking per-file diarization speakers via speaker 
        embeddings \cite{Wang2023} using a 0.6 cosine-similarity merge threshold. 
        As shown in Fig.~\ref{fig:duration_distribution}, utterance durations 
        are concentrated in the 10--20-sec range, with a mean of 
        $13.87 \pm 6.32$~s closely matching our target Gaussian sampling 
        distribution ($\mu=12~\mathrm{s}$) described in 
        Section~\ref{subsubsec:prosody_aware}---confirming that our prosody-aware 
        segmentation strategy achieves durations well-suited for long-context TTS 
        modeling \cite{makarov2022simpleeffectivemultisentencetts}.

        \begin{table}[!t]
            \centering
            \small
            \caption{Summary statistics of the final PersianVox dataset.}
            \label{tab:corpus_stats}
            \begin{tabular}{lc}
                \hline
                \textbf{Statistic} & \textbf{Value} \\ \hline
                Total duration & 2,408.67 h \\
                Number of utterances & 625,192 \\
                Unique speakers & 3,248 \\
                Utterance duration & $13.87 \pm 6.32$~s \\
                Vocabulary size & 212,216 \\
    
                \hline
            \end{tabular}
        \end{table}
    	
	\subsubsection{Effect of Speech Restoration}
\begin{figure}[!t]
        	\centering
        	\includegraphics[width=0.85\linewidth]{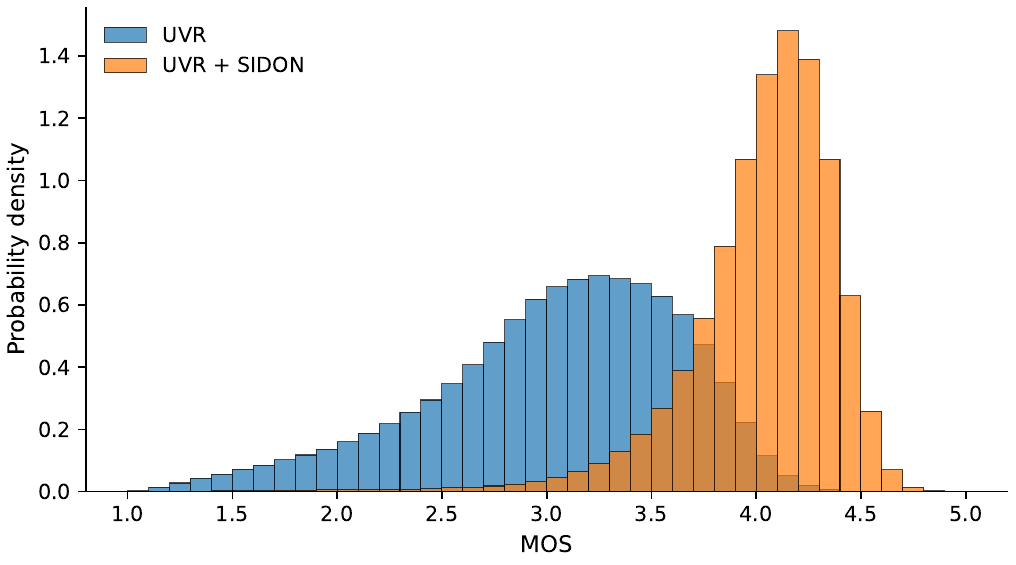}
        	\caption{SCOREQ MOS distributions comparing utterances processed with 
        	UVR only versus UVR followed by 
        	SIDON speech restoration.}
        	\label{fig:restoration_histogram}
        \end{figure}

        To quantify the benefit of speech restoration, we compare SCOREQ MOS scores between segments processed with only UVR music removal (UVR-only) and those additionally restored with per-utterance SIDON after segmentation (UVR + SIDON). UVR-only segments achieve a mean SCOREQ score of 3.04 (std 0.6), while UVR + SIDON segments reach 4.03 (std 0.36), showing improved perceptual quality and reduced variance. As shown in Fig.~\ref{fig:restoration_histogram}, the restored segments are shifted toward higher quality scores, with nearly all exceeding SCOREQ 3.5, indicating that SIDON effectively reduces residual degradations such as reverberation and codec artifacts after music removal.
    
	\subsubsection{Prosody-Aware Duration Distribution}
    	Fig.~\ref{fig:duration_distribution} demonstrates the empirical impact of our prosody-aware approach compared to the Emilia pipeline (blue). Our method (orange) successfully shifts the distribution rightward toward a more balanced, Gaussian-like curve. This structural improvement increases the proportion of longer segments, equipping TTS models with the necessary context for superior naturalness and expressivity.
	
\section{SQA Benchmarking and MOS Release} \label{section:sqa_benchmark}
    To validate our quality control stage and address the scarcity of Persian speech quality resources, we curated a 1.6-hour subset (564 utterances) of PersianVox and collected MOS annotations from native speakers. We release both audio and MOS labels to support future evaluation. We benchmarked five SQA methods to select the most reliable filter for our pipeline. As shown in Table~\ref{tab:sqa_results}, SCOREQ \cite{ragano2025scoreqspeechqualityassessment} achieves the highest Pearson (PLCC) and Spearman (SRCC) correlations, outperforming the industry baseline DNSMOS \cite{reddy2022dnsmosp835nonintrusiveperceptual}.
	
    \begin{table}[!t]
        \centering
        \small
        \caption{Performance of SQA methods on the PersianVox MOS subset ($N=564$).}
        \label{tab:sqa_results}
        \begin{tabular}{lccc}
            \hline
            \textbf{Method} & \textbf{PLCC} & \textbf{SRCC} & \textbf{RMSE} \\ \hline
            DNSMOS \cite{reddy2022dnsmosp835nonintrusiveperceptual} & 0.5590 & 0.5182 & 0.6779 \\
            NISQA \cite{Mittag_2021} & 0.4786 & 0.4703 & 0.7179 \\
            \textbf{SCOREQ \cite{ragano2025scoreqspeechqualityassessment}} & \textbf{0.6658} & \textbf{0.6271} & \textbf{0.6100} \\
            VQScore \cite{fu2024selfsupervisedspeechqualityestimation} & 0.3826 & 0.3332 & 0.7554 \\
            WhiSQA \cite{close2025whisqanonintrusivespeechquality} & 0.6003 & 0.5704 & 0.6539 \\
            \hline
        \end{tabular}
    \end{table}
	
\section{TTS Model Training}
	To validate PersianVox, we adapt LinaSpeech \cite{lemerle2025linaspeechgatedlinearattention} by replacing its WavTokenizer with Vevo's flow-matching acoustic decoder \cite{zhang2025vevocontrollablezeroshotvoice}, which generalizes seamlessly to TTS and achieves significantly higher perceptual quality for Persian than standard codec-based alternatives.
	
    \subsection{TTS Evaluation}
    To evaluate the synthesized speech, we compute the WER and CER using
    transcriptions from our 600M-parameter Parakeet RNN-T model. Speaker
    preservation is measured via Speaker Embedding Cosine Similarity (SECS) using CAM++ \cite{wang2023camfastefficientnetwork}
    against the reference prompts. For perceptual quality, we report the
    objective MOS using SCOREQ, which we identified as the most reliable
    quality metric for Persian (Section~\ref{section:sqa_benchmark}). As
    summarized in Table~\ref{tab:tts_results}, the model demonstrates strong
    intelligibility, speaker similarity, and naturalness, validating PersianVox
    as an effective TTS training resource.

    \begin{table}[!ht]
        \centering
        \small
        \caption{Objective evaluation of the TTS model trained on PersianVox.}
        \label{tab:tts_results}
        \setlength{\tabcolsep}{3pt}
        \begin{tabular}{lcccc}
            \hline
            \textbf{Model} & \textbf{WER} $\downarrow$ &
            \textbf{CER} $\downarrow$ & \textbf{SECS} $\uparrow$ &
            \textbf{MOS} $\uparrow$ \\ \hline
            Reference & -- & -- & -- & 4.31 \\
            Synthesized & 2.1 & 0.3 & 79.8 & 4.27 \\
            \hline
        \end{tabular}
    \end{table}
    \FloatBarrier

\section{Conclusion}
    We present PersianVox, a scalable framework for constructing high-quality speech datasets from in-the-wild audio. The proposed prosody-aware segmentation strategy preserves linguistic completeness and provides longer contextual units for TTS training, while dual-ASR agreement improves transcription reliability without requiring ground-truth labels. Our SQA evaluation further identifies SCOREQ as an effective quality filter for Persian speech. The resulting 2,400-hour corpus and human-annotated benchmark provide a strong foundation for Persian speech synthesis and offer a practical framework for dataset creation in other low-resource languages.
	
\bibliographystyle{IEEEtran}
\bibliography{mybib}
\end{document}